\documentclass{INTERSPEECH2023}

\usepackage{graphicx}
\usepackage{subfigure}
\usepackage{booktabs}
\usepackage{cite}
\usepackage{microtype}
\usepackage{xurl}
\usepackage{amsmath}
\usepackage{amssymb}
\usepackage{mathtools}
\usepackage{amsthm}
\usepackage{textcmds}
\usepackage{color}
\usepackage[flushmargin]{footmisc}
\usepackage{arydshln}
\usepackage{comment}

\definecolor{azure(colorwheel)}{rgb}{0.0, 0.5, 1.0}
\definecolor{nicegreen}{rgb}{0.0, 0.7, 0.1}
\definecolor{CuGray}{gray}{0.9}
\definecolor{pink}{cmyk}{0, 0.7808, 0.4429, 0.1412}
\definecolor{amethyst}{rgb}{0.6, 0.4, 0.8}
\definecolor{black}{rgb}{0.0, 0.0, 0.0}
\definecolor{purple}{rgb}{0.6275, 0.0275, 0.6706}

\definecolor{steelblue}{rgb}{0.27, 0.51, 0.7}
\definecolor{brickred}{rgb}{0.8, 0.25, 0.33}

\definecolor{customgray}{rgb}{0.9, 0.9, 0.9}
\renewcommand{\paragraph}[1]{\vspace{1mm}\noindent\textbf{#1.}\,\,\,}

\usepackage{enumitem}
\usepackage{multirow}

\usepackage[breaklinks,colorlinks,citecolor=nicegreen,linkcolor=red,bookmarks=false,pdfauthor={Author name(s) withheld},pdftitle={Submitted to INTERSPEECH}]{hyperref}
\usepackage[capitalize,noabbrev]{cleveref}

\usepackage{enumitem}
\usepackage{multirow}
\usepackage{wrapfig}
\usepackage{colortbl}
\usepackage[normalem]{ulem}
\usepackage{comment}

\setlist[itemize]{align=parleft,left=0pt,topsep=1mm,itemsep=0mm,parsep=1mm}

\definecolor{azure(colorwheel)}{rgb}{0.0, 0.5, 1.0}
\definecolor{nicegreen}{rgb}{0.0, 0.7, 0.1}
\definecolor{yw}{rgb}{0.01176, 0.5490, 0.5490}
\definecolor{ashblue}{rgb}{0.36, 0.54, 0.66}
\definecolor{ashgrey}{rgb}{0.7, 0.75, 0.71}
\definecolor{applegreen}{rgb}{0.55, 0.71, 0.0}
\definecolor{greenyellow}{rgb}{0.68, 1.0, 0.18}
\definecolor{junebud}{rgb}{0.74, 0.85, 0.34}
\definecolor{kellygreen}{rgb}{0.3, 0.73, 0.09}
\definecolor{ywg}{rgb}{0.9960, 0.8984, 0.5859}
\definecolor{jy}{rgb}{0.58, 0, 0.827}
\definecolor{cornellred}{rgb}{0.7, 0.11, 0.11}
\definecolor{darkcyan}{rgb}{0.0, 0.55, 0.55}
\definecolor{CuGray}{gray}{0.9}
\definecolor{airforceblue}{rgb}{0.36, 0.54, 0.66}
\definecolor{rev}{rgb}{0.784, 0.003, 0.313}
\definecolor{pink}{cmyk}{0, 0.7808, 0.4429, 0.1412}
\definecolor{amethyst}{rgb}{0.6, 0.4, 0.8}
\definecolor{black}{rgb}{0.0, 0.0, 0.0}
\definecolor{tb3_yellow}{rgb}{0.996, 1.0, 0.6}
\definecolor{tb3_orange}{rgb}{0.980, 0.8, 0.604}
\definecolor{tb3_red}{rgb}{0.972, 0.6, 0.6}
\definecolor{dimgray}{rgb}{0.41, 0.41, 0.41}
\definecolor{brickred}{rgb}{0.8, 0.25, 0.33}
\definecolor{bleudefrance}{rgb}{0.19, 0.55, 0.91}
\definecolor{blue(ncs)}{rgb}{0.265, 0.445, 0.765}
\definecolor{green(ncs)}{rgb}{0.0, 0.62, 0.42}

\newcolumntype{g}{>{\columncolor{CuGray}}c}
\newcolumntype{z}{>{\columncolor{CuGray}}l}

\renewcommand{\paragraph}[1]{\vspace{0.5mm}\noindent\textbf{#1.}\,\,}

\usepackage{xspace}

\makeatletter
\def\@fnsymbol#1{\ensuremath{\ifcase#1\or *\or \dagger\or \ddagger\or
   \mathsection\or \mathparagraph\or \|\or **\or \dagger\dagger
   \or \ddagger\ddagger \else\@ctrerr\fi}}
\makeatother

\def\onedot{.\@\xspace}
\def\eg{\emph{e.g}\onedot} 
\def\ie{\emph{i.e}\onedot}

\newcommand{\be}{\begin{eqnarray}}
\newcommand{\ee}{\end{eqnarray}}
\newcommand{\bee}{\begin{eqnarray*}}
\newcommand{\eee}{\end{eqnarray*}}

\newcommand{\matrixb}{\left[ \begin{array}}
\newcommand{\matrixe}{\end{array} \right]}

\usepackage{mathtools}

\DeclarePairedDelimiterX{\inp}[2]{\langle}{\rangle}{#1, #2}

\interspeechcameraready

\title{Automatic Tuning of Loss Trade-offs without Hyper-parameter Search in End-to-End Zero-Shot Speech Synthesis}
\name{Seongyeon Park$^1$, Bohyung Kim$^1$, Tae-hyun Oh$^{2,3}$\thanks{\textbf{Acknowledgements}: T.-H. Oh was supported by 
IITP
grants funded by the Korea government(MSIT) (No.2021-0-02068, Artificial Intelligence Innovation Hub; No.2022-0-00124, Development of Artificial Intelligence Technology for Self-Improving Competency-Aware Learning Capabilities).}}
\address{
  $^1$CNAI, Korea\\
  $^2$Institute for Convergence Research and Education in Advanced Technology, Yonsei University, Korea\\
  $^3$Dept. of EE and GSAI, POSTECH, Korea.}
\email{eaglesy221@snu.ac.kr}

\begin{document}

\maketitle
 
\begin{abstract}
Recently, zero-shot TTS and VC methods have gained attention due to their practicality of being able to generate voices even unseen during training.
Among these methods, zero-shot modifications of the VITS model have shown superior performance, while having useful properties inherited from VITS.
However, the performance of VITS and VITS-based zero-shot models vary dramatically depending on how the losses are balanced.
This can be problematic, as it requires a burdensome procedure of tuning loss balance hyper-parameters to find the optimal balance.
In this work, we propose a novel framework that finds this optimum without search, by inducing the 
decoder of VITS-based models to its full reconstruction ability.
With our framework, we show superior performance compared to baselines in zero-shot TTS and VC, achieving state-of-the-art performance.
Furthermore, we show the 
robustness
of our framework in various settings.
We provide an explanation for the results in the discussion.
\end{abstract}

\section{Introduction}
The recent advance of neural networks has enhanced the quality of speech synthesis models in the areas of Text-to-Speech (TTS) and Voice Conversion (VC), resulting in the production of realistic and naturally sounding speech. 
Among multiple categories of TTS and VC, zero-shot approaches, \eg, \cite{jia2018transfer, cooper2020zero, choi20c_interspeech, min2021meta,chou19_interspeech,qian2019autovc,wu20p_interspeech,wu2020one,wang21n_interspeech,yang22f_interspeech,lian22_interspeech}, have obtained much attention.
These approaches are practical, because they can synthesize speech of given voices even unseen during training.

While models for zero-shot VC and TTS were developed separately, Glow-TTS~\cite{kim2020glow} enables VC and multi-speaker TTS with a single model, utilizing the invertibility of Normalizing Flows (NFs)~\cite{rezende2015variational}.
Building upon this,
VITS~\cite{kim2021conditional} extends it 
with Variational Autoencoder (VAE)~\cite{kingma2013auto} structure where its decoder is HiFi-GAN~\cite{kong2020hifi}.
By jointly training its components, VITS is not only practical in terms that it works in a single stage (\ie, can predict waveform directly), but can also produce ground truth level fidelity speech
of multiple speakers.

By virtue of these favorable properties,
there have been attempts~\cite{casanova21b_interspeech, cong21_interspeech, lei22_interspeech, casanova2022yourtts} to extend VITS or Glow-TTS to zero-shot VC and TTS.
They 
use speaker encoders instead of the speaker embedding table in VITS or Glow-TTS, so that the models can extract and synthesize diverse unseen voices in the inference phase, \ie, enabling zero-shot.
While extending the useful properties of Glow-TTS and/or VITS, these zero-shot models reported superior performance over previous zero-shot VC and TTS models.
In particular, VITS-based models~\cite{casanova2022yourtts, lei22_interspeech} show superior performance compared to Glow-TTS-based models~\cite{casanova21b_interspeech, cong21_interspeech}, as the models can internally optimize the latent representations instead of using human-designed features such as mel-spectrograms.

\begin{figure}[t!]
\centering

\begin{tabular}{c}
    \includegraphics[width=\linewidth]{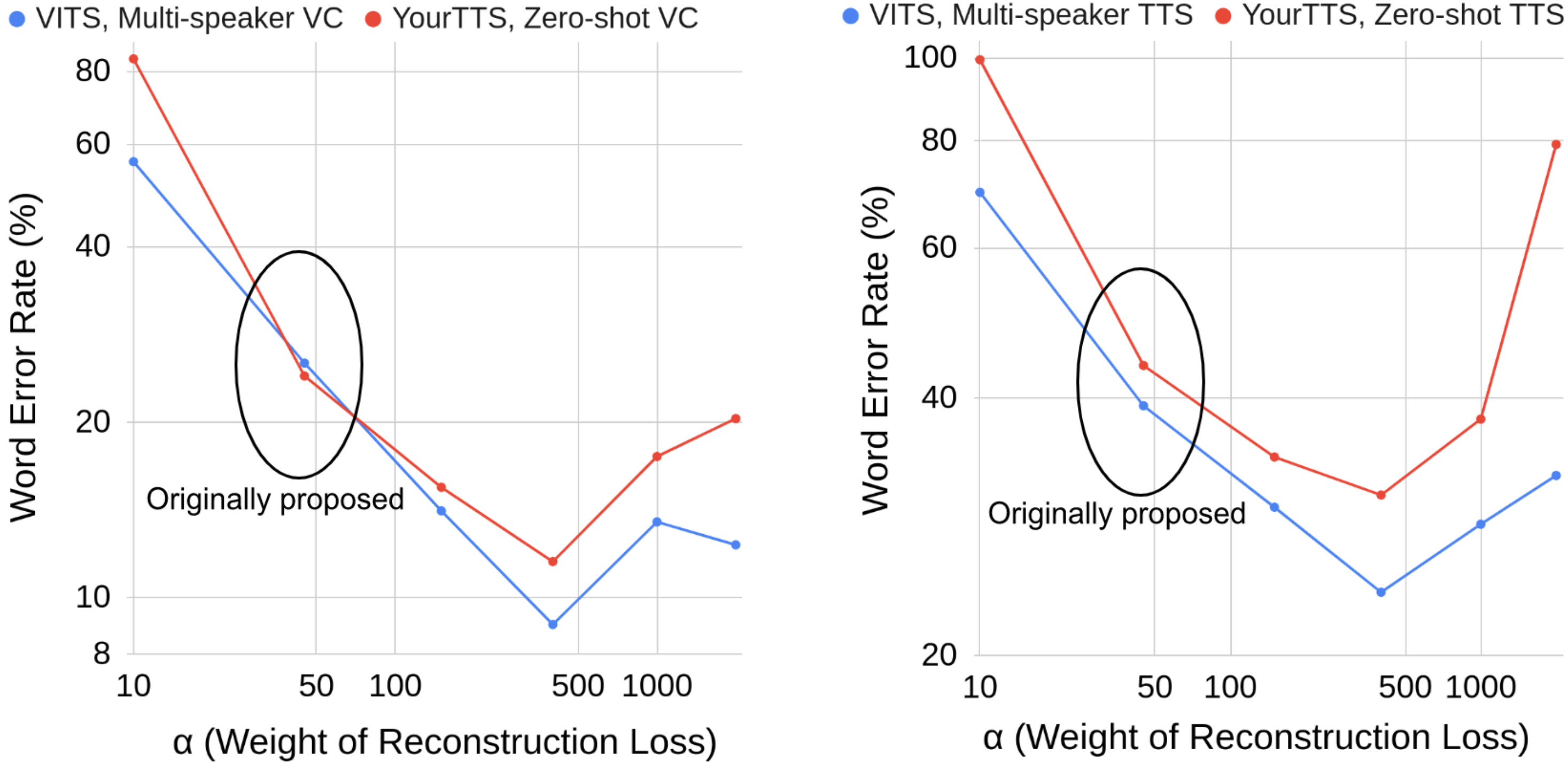}
\end{tabular}

\caption{Word Error Rate of speech synthesized through VC or TTS, from VITS and YourTTS~\cite{casanova2022yourtts} according to the loss balance hyper-parameter $\alpha$ (the loss weight parameter of the reconstruction loss).
Both axes are in log scale.
The WER has a noticeable variance depending on the balance hyper-parameter.
Interestingly, the results obtained with the parameter suggested by the original authors show a notable gap from the best points we found.
These graphs suggest that there is likely to exist an optimal value of the hyper-parameter $\alpha$.
One might exhaustively tune balance hyper-parameters for better performance.
In this work, we propose a framework that alleviates the exhaustive hyper-parameter tuning process, and instead enables a near-optimal trade-off selection without search.}
\label{fig:intro_tradeoff}
\end{figure}

However, we found that VITS and zero-shot modifications of VITS 
tend to be sensitive to 
how losses are balanced.
Specifically, performance depending on the balance hyper-parameter for reconstruction loss can be seen in \cref{fig:intro_tradeoff}.
A naive approach to finding the optimal point would involve searching and tuning the loss balance hyper-parameters.
In this work, we propose a framework that alleviates the exhaustive 
tuning process and instead enables a near-optimal trade-off selection without search.

Our proposed method is built on our hypothesis that the converged reconstruction loss value of separately training the HiFi-GAN vocoder would provide a favorable reference 
for the reconstruction loss in training VITS-based models.
More specifically, as a preliminary task,
we first train just 
HiFi-GAN as a mel-spectrogram-to-waveform vocoder.
Denoting the converged reconstruction loss value 
as $\varepsilon^*$, we obtain this value 
from the preliminary task.
Then, we train VITS-based models for the main speech synthesis task, such that its reconstruction loss is converged at $\varepsilon^*$, which would be sufficient to get high fidelity audios even in the main task.
We realize this learning process 
by the Modified Differential Multiplier Method (MDMM) 
\cite{NIPS1987_a87ff679}.

With this method, our framework solves the pre-mentioned trade-off problem in VITS-based models without hyper-parameter tuning.
We experimentally show the superior performance achieved by this framework in zero-shot VC and TTS, achieving state-of-the-art performance.
We also show its robustness across various model and audio configurations.
We further explain the results in the discussion.
Source code and audio samples are available at: \url{https://github.com/cnaigithub/Auto_Tuning_Zeroshot_TTS_and_VC}

\section{Method}\label{Method}
In this section, we first describe the base models used in this work.
We explain VITS~\cite{kim2021conditional}, the state-of-the-art model for multi-speaker speech synthesis, and introduce a simple variant of VITS that enables zero-shot synthesis.
Then we describe our framework and how to apply it to the base models.

\subsection{Base Model for Speech Synthesis}


VITS basically integrates Glow-TTS~\cite{kim2020glow} and HiFi-GAN~\cite{kong2020hifi}.

Glow-TTS is a Normalizing Flow (NF)~\cite{rezende2015variational}-based TTS model, which uses 
the monotonic alignment search 
algorithm to obtain a hard attention alignment between input text and target mel-spectrogram timesteps.
The original Glow-TTS is designed to conduct text-to-mel-spectrogram conversion, \ie, TTS.

HiFi-GAN is a waveform synthesizer, which typically decodes mel-spectrograms into speech waveforms via adversarial training.
Instead of mel-spectrograms as input,
the HiFi-GAN used as the decoder in VITS decodes latent features obtained from Glow-TTS and directly converts to waveforms, which enables achieving high fidelity audio synthesis.

With these modules, VITS integrates them and adds the 
VAE
structure.
The latent representation of this VAE follows a posterior distribution, encoded from linear spectrograms of speech audio.
Also, a text-conditioned prior distribution
is produced by the Glow-TTS module.
These posterior and prior are regularized by the KL divergence during training.
Noticeably, the invertibility of NF in VITS allows latent features to go forward and backward, which enables performing both VC and TTS in a single model.
In addition, to achieve multi-speaker speech synthesis, VITS adopts a trainable speaker embedding table. 
The speaker embeddings are injected to condition the posterior encoder, the NF layer, and the HiFi-GAN decoder.

\paragraph{Training of VITS}
VITS employs adversarial training in the waveform domain by training 
both a generator and discriminator network.
The generator and discriminator are optimized by the following respective losses,
\begin{equation}\label{VITS loss}
\begin{aligned}
    &L_{g} = \alpha L_{recon} + L_{KL} + L_{dur} + L_{adv}(G; D) + L_{fm}(G; D), \\
    &L_{d} = L_{adv}(D; G),
\end{aligned}
\end{equation}
where the reconstruction loss term $L_{recon}$ is defined by an $L_1$ loss between the mel-spectrogram of the generated waveform and ground truth waveform, $L_{adv}$ is the adversarial loss, and $L_{KL}$ is the KL divergence between the posterior obtained from audio and the prior obtained from text.
For detail on the other losses $L_{dur}$ and $ L_{fm}$, refer to the literature~\cite{kim2021conditional}.
The balance hyper-parameter $\alpha$ weights the reconstruction loss against other losses, and is the same one used in \cref{fig:intro_tradeoff}.
As the model outputs waveform, a differentiable STFT and linear projection to the mel-scale is required to back-propagate from $L_{recon}$~\cite{kim2021conditional}.

Note that, when training HiFi-GAN as a mel-spectrogram-to-waveform vocoder, it uses the same losses for its generator and discriminator as \cref{VITS loss}, except for excluding $L_{KL}$ and $L_{dur}$.
This enables a direct comparison of $L_{recon}$ values in HiFi-GAN training with the one in VITS training despite noticeable differences in modeling and tasks.

\paragraph{Zero-shot Modification of VITS}
\label{Zero-shot VITS}
We modify and extend the original multi-speaker VITS model with simple changes in order to enable end-to-end zero-shot speech synthesis.
To prevent performance drop on non-English datasets, we use a jointly trained speaker encoder instead of a pre-trained one.
We replace the speaker embedding table of VITS with a speaker encoder consisting of convolutional layers.
The architecture of the speaker encoder is similar to the one in \cite{lei22_interspeech}, except that our speaker encoder takes linear spectrograms as the input source.
Second, we do not condition the posterior encoder by speaker embedding.
This is meant to prevent performance drop on voice conversion with unseen source speakers.
Third, we use 10~transformer blocks for the text encoder following YourTTS~\cite{casanova2022yourtts}.
Fourth, we use the Deterministic Duration Predictor (DDP) instead of the Stochastic Duration Predictor (SDP).
This is due to reports on the instability of SDP in \cite{casanova2022yourtts, cho22_interspeech}.
For simplicity, we call this modified model \textit{Zero-shot VITS}.
Through experiments, we show that this model shows comparable performance to YourTTS~\cite{casanova2022yourtts} (a baseline zero-shot modification of VITS) without our framework applied, but outperforms it when our framework is applied.

\subsection{
Loss Value of Separate Vocoder Training as a 
Guide}
As shared in all VAE-based methods, both VITS and zero-shot modifications of VITS show trade-offs between their reconstruction loss and the other losses.
By changing the balance ratio of these losses, these models converge on different Pareto equilibria.
As mentioned before, model performance varies drastically depending on which Pareto equilibrium it converged (\cref{fig:intro_tradeoff}).
A vanilla method to find the optimum would be to conduct exhaustive iterative experiments with different balance parameters, which is computationally burdensome.
Instead of directly attempting different balance hyper-parameters $\alpha$, we tune a specific target value of $L_{recon}$ to find an optimal balance.

The motivation of this approach is that the decoder of a VITS-based model is a HiFi-GAN; thus, if HiFi-GAN works well for its own vocoder task, the level of the converged value of $L_{recon}$ may indicate the desired quality and may be transferred to other tasks.
To test this hypothesis, we separately train HiFi-GAN for a vocoder task alone, we call a preliminary task.
Then, we obtain the converged $L_{recon}$ value, denoted as $\varepsilon^*$, and do not use any other results, including the trained vocoder model from the preliminary task.
In our framework, our idea is to bring and use $\varepsilon^*$ as the target value to converge $L_{recon}$ in VITS training.
That is, we propose to use this particular $L_{recon}$ value as the desired point for a trade-off between $L_{recon}$ and other losses.

\subsection{Constraint Optimization of VITS to 
a Specific $L_{recon}$}

\paragraph{Constrained Optimization for Neural Networks}
Platt and Barr~\cite{NIPS1987_a87ff679} introduces methods for constrained optimization on neural networks.
We briefly explain the Modified Differential Multiplier Method (MDMM)
proposed in \cite{NIPS1987_a87ff679}.
Consider a neural network, whose parameters are $\theta$.
Suppose we would like to solve the following optimization problem:
\begin{equation}\label{Eq1}
\begin{split}
    \min\nolimits_{\theta}F(\theta) 
    \quad\textrm{s.t.}\quad
    \ G(\theta)=0
\end{split}
\end{equation}
To optimize such problem, the Lagrange multiplier method can be used.
To be more specific, the solution $\theta$ of \cref{Eq1} should be a critical point of the following Lagrangian function:
\begin{equation}\label{Eq2}
\begin{split}
    \mathcal{L}(\theta, \lambda)=F(\theta) + \lambda G(\theta)
\end{split}
\end{equation}
for some Lagrange multiplier $\lambda$.
Note that the solution is also a critical point for $\lambda$, 
as $\frac{\partial \mathcal{L}}{\partial \lambda} = G(\theta)= 0$.
The Lagrange multiplier $\lambda$ can be regarded as an additional variable and can be updated via gradient descent or ascent.
The MDMM method adds a penalty term for the constraint as:
\begin{equation}\label{Eq3}
\begin{split}
    \mathcal{L}(\theta, \lambda)=F(\theta) + \lambda G(\theta) + \tfrac{c}{2} G(\theta)^2,
\end{split}
\end{equation}
where $c$ is the damping constant parameter.
The MDMM updates $\theta$ with gradient descent and $\lambda$ with gradient ascent over $\mathcal{L}$.
Please refer to more details.
\cite{NIPS1987_a87ff679}.

\paragraph{Selecting a VITS' Trade-Off Point with MDMM}
We can select a trade-off point of VITS 
by enforcing the reconstruction loss $L_{recon}$ to a user-chosen constant value $\varepsilon$.
To be more specific, we define $G(\theta)=L_{recon}(\theta)-\varepsilon$.
The loss $F(\theta)$ is set to be the sum of all the remaining loss terms. 
Then, we can update $\theta$ and $\lambda$ according to the following gradients:
\begin{equation}\label{Eq4}
\begin{split}
    &\tfrac{\partial \mathcal{L}}{\partial \lambda} = G(\theta)=L_{recon}(\theta)-\varepsilon \\
    &\tfrac{\partial \mathcal{L}}{\partial \theta} = \tfrac{\partial F}{\partial \theta}+\lambda\tfrac{\partial L_{recon}}{\partial \theta}+c(L_{recon}-\varepsilon)\tfrac{\partial L_{recon}}{\partial \theta}
\end{split}
\end{equation}

By this optimization, the model converges on the specified Pareto equilibrium, where $L_{recon}$ is equal to $\varepsilon$.
In our proposed framework, 
we set 
$\varepsilon=\varepsilon^*$, but for the sake of proving its empirical optimality, we also report model performance on nearby values, specifically $\varepsilon=\varepsilon^* \pm 0.1$.


\section{Experiments, Results and Discussion}
\subsection{Experiment
Settings and Configurations}
\paragraph{Evaluation}
We use Word Error Rate (WER) / Character Error Rate (CER) of the synthesized speech and Resemblyzer Embedding Cosine Similarity (RECS)~\cite{wan2018generalized}\footnote{https://github.com/resemble-ai/Resemblyzer} between the target speech and synthesized speech.
We use a pre-trained ASR model\footnote{https://huggingface.co/facebook/hubert-large-ls960-ft} for obtaining the WER / CER between synthesized speech and the ground truth transcript.

\paragraph{Datasets}
We either use the VCTK~\cite{veaux2016superseded} dataset or a multilingual dataset mostly composed of public data\footnote{The dataset consists of 15 different languages, mostly publicly available data.
English~\cite{veaux2016superseded, zen19_interspeech}, Dutch / French / German / Italian / Polish / Portuguese / Spanish~\cite{pratap20_interspeech, mailabs}, Thai / Cantonese~\cite{ardila-EtAl:2020:LREC}, Korean~\cite{park2018kss}, Japanese~\cite{takamichi2019jvs, sonobe2017jsut}, Mandarin~\cite{shi2020aishell}, Vietnamese~\cite{luong-vu-2016-non}, and Arabic~\cite{tunisian}.
We also add some proprietary data containing 4 Korean speakers, which take up less than 0.5\% of the entire training set.
We exclude any data longer than 13~seconds, and trim silences with librosa~\cite{mcfee2015librosa}.

}.
When using the VCTK dataset, we hold out 4 male and 4 female speakers from its training set and use them for evaluation on unseen speakers.
For evaluation on seen speakers, we randomly hold out 128 utterances from the remaining 
 training set.
The rest not used for evaluation is used for training.
Evaluation for models trained with the multilingual dataset was conducted using the \textit{test-clean} split of LibriTTS~\cite{zen19_interspeech}, consisting of speakers unseen during training.
The rest is used for training.\footnote{For training \textit{Zero-shot VITS} with the multilingual dataset, we add a language embedding lookup table and condition the model on language embeddings, by following the multilingual training of YourTTS.
}
We use International Phonetic Alphabet as transcription, following VITS~\cite{kim2021conditional}.



\paragraph{Detail Configurations}
We downsampled audio to 16kHz, and used linear/mel spectrograms with a hop size of 320 and a window size of 1280. 
The sampling rate and hop size were selected according to the input and output frequencies of WavLM~\cite{chen2022wavlm} features, which are used in one of the baselines~\cite{lian22_interspeech}.
We trained each model with a batch size of 32 for 500k steps.
We use the AdamW optimizer for our methods, with a learning rate of 2e-4 and a weight decay of 0.01.
For the preliminary task, we train HiFi-GAN for 500k steps, as improvement in perceptual quality is small afterward.
We used the optimizers and learning rates specified in the original works for the other methods.

\subsection{Effectiveness of Our Framework}
To see the effectiveness of our framework,
we apply our framework to VITS and zero-shot modifications of VITS, and compare them with 
the state-of-the-arts in zero-shot VC / TTS on VCTK~\cite{veaux2016superseded}.
We denote this experiment as \texttt{Exp1}.

\paragraph{Compared Methods}
YourTTS~\cite{casanova2022yourtts} is a VITS-based model, that uses a language embedding table to be trained as a multilingual TTS model.
It uses a pre-trained speaker encoder to enable zero-shot synthesis, and uses the speaker consistency loss
to improve the similarity of the synthesized and reference speeches.
Its VITS-based structure enables both TTS and VC.
C-DSVAE~\cite{lian22_interspeech} is a VAE-based VC model, and has separate encoders for 
speaker information and content.
Due to its fast convergence, we train C-DSVAE for 150k steps.
We refer the readers to the original papers~\cite{casanova2022yourtts, lian22_interspeech} for more details.
C-DSVAE is a two-stage model, which predicts the mel-spectrogram and thus needs an external vocoder to produce waveform.
We use the separate HiFi-GAN vocoder~\cite{kong2020hifi} trained on  VCTK for 500k steps with our configurations.

\begin{table*}[t] 
\caption{Comparisons between the prior arts and the methods  
with our framework.
\textit{Zero-shot VITS} combined with our framework shows favorable performance compared to the prior arts without our framework in zero-shot VC and TTS.
Further, the performance of VITS and YourTTS is improved significantly when our framework is applied, especially in the VC task.
The best results 
within the same setting are marked in bold.
\vspace{-2mm}}
\label{tab:vctk_wer_recs}
\centering
\resizebox{\linewidth}{!}{
\begin{tabular}{c|cccccccccc}
\hline
\multirow{2}{*}{Method}                         & \multicolumn{2}{c}{VC (Seen-to-seen)}              & \multicolumn{2}{c}{VC (Unseen-to-seen)}                                 & \multicolumn{2}{c|}{VC (Unseen-to-unseen)}                              & \multicolumn{2}{c}{TTS (Seen)}                     & \multicolumn{2}{c|}{TTS (Unseen)}                  \\ \cline{2-11} 
                                                & WER/CER (\%) ($\downarrow$)          & RECS ($\uparrow$)                      & \multicolumn{1}{c|}{WER/CER (\%)}          & RECS                       & WER/CER (\%)          & \multicolumn{1}{c|}{RECS}                       & WER/CER (\%)          & RECS                       & WER/CER (\%)          & RECS                       \\ \hline
GT (VCTK)                                       & \multicolumn{10}{c|}{6.86 / 2.24 \quad\quad 0.803 $\pm$ 0.008}                                                                                                                                                                                                                                                   \\ \hline
VITS~\cite{kim2021conditional}                  & 16.03 / 7.12          & 0.792 $\pm$ 0.010          & \multicolumn{1}{c|}{$\times$}              & $\times$                   & $\times$              & \multicolumn{1}{c|}{$\times$}                   & 29.87 / 16.21         & 0.801 $\pm$ 0.001          & $\times$              & $\times$                   \\
C-DSVAE~\cite{lian22_interspeech}               & 44.05 / 24.80         & 0.709 $\pm$ 0.012          & \multicolumn{1}{c|}{46.89 / 27.20}         & 0.693 $\pm$ 0.009          & 45.23 / 25.69         & \multicolumn{1}{c|}{0.614 $\pm$ 0.010}          & $\times$              & $\times$                   & $\times$              & $\times$                   \\
YourTTS~\cite{casanova2022yourtts}              & 16.78 / 7.94          & 0.814 $\pm$ 0.010          & \multicolumn{1}{c|}{26.61 / 13.40}         & 0.791 $\pm$ 0.007          & 24.04 / 11.90         & \multicolumn{1}{c|}{0.734 $\pm$ 0.009}          & 37.64 / 19.83         & 0.817 $\pm$ 0.010          & 36.28 / 20.27         & 0.751 $\pm$ 0.009          \\
\textit{Zero-shot VITS} (\cref{Zero-shot VITS}) & 20.79 / 10.53         & \textbf{0.815 $\pm$ 0.010} & \multicolumn{1}{c|}{29.53 / 15.54}         & \textbf{0.804 $\pm$ 0.007} & 26.53 / 13.23         & \multicolumn{1}{c|}{\textbf{0.744 $\pm$ 0.008}} & 28.06 / 13.87         & \textbf{0.818 $\pm$ 0.010} & 27.38 / 14.38         & 0.747 $\pm$ 0.009          \\
VITS with our framework                         & \textbf{10.52 / 4.24} & 0.786 $\pm$ 0.011          & \multicolumn{1}{c|}{$\times$}              & $\times$                   & $\times$              & \multicolumn{1}{c|}{$\times$}                   & 26.22 / 13.22         & 0.794 $\pm$ 0.010          & $\times$              & $\times$                   \\
YourTTS with our framework                      & 11.34 / 5.16          & 0.811 $\pm$ 0.010          & \multicolumn{1}{c|}{\textbf{14.19 / 6.60}} & 0.768 $\pm$ 0.008          & \textbf{13.72 / 6.09} & \multicolumn{1}{c|}{0.730 $\pm$ 0.009}          & 33.27 / 17.16         & 0.812 $\pm$ 0.010          & 32.34 / 17.26         & 0.755 $\pm$ 0.008          \\
\textit{Zero-shot VITS} with our framework      & 12.18 / 5.94          & 0.800 $\pm$ 0.010          & \multicolumn{1}{c|}{15.07 / 7.20}          & 0.782 $\pm$ 0.007          & 14.46 / 6.69          & \multicolumn{1}{c|}{0.740 $\pm$ 0.008}          & \textbf{19.25 / 9.14} & 0.814 $\pm$ 0.010          & \textbf{19.36 / 9.04} & \textbf{0.762 $\pm$ 0.009} \\ \hline
\end{tabular}
}\vspace{-2mm}
\end{table*}

\paragraph{Results}
The HiFi-GAN trained with our settings converged to $L_{recon} = 0.25$.
Thus we use $\varepsilon = \varepsilon^* = 0.25$ for the MDMM optimization framework in \texttt{Exp1}\footnote{The VITS-based models used in \texttt{Exp1}, \ie VITS, YourTTS, and \textit{Zero-shot VITS}, have identical decoders, and thus share the same $\varepsilon^*$ value.}.
As shown
in \cref{tab:vctk_wer_recs}, \textit{Zero-shot VITS} with 
our framework shows favorable performance in all settings of VC and TTS compared to previously proposed state-of-the-art zero-shot models.
Also, VITS and YourTTS' performance is significantly improved by our framework.



\subsection{Robustness of Our Framework}
We experiment our framework with different models and data configurations to show the  robustness of our method.
For each experiment, we first train HiFi-GAN vocoders with different conditions to obtain $\varepsilon^*$.
Then, we train VITS-based models following each setting
with $\varepsilon{=}\varepsilon^*$ and nearby values $\varepsilon{=}\varepsilon^*{\pm}0.1$.
Unless specified otherwise, VCTK is used as the default dataset,
and \textit{Zero-shot VITS} as the default model.
We train models for 300k steps as metric improvements were small afterward.
{\footnotesize
\begin{itemize}
\item{\texttt{Exp2} (Different dataset): Use the multilingual dataset.}
\item{\texttt{Exp3} (Decoder with less reconstruction ablility): Reduce the internal channel numbers of HiFi-GAN and \textit{Zero-shot VITS}' decoder by a factor of 8.}
\item{\texttt{Exp4} (Different model with same decoder): Use YourTTS, which has the same decoder as \textit{Zero-shot VITS}.}
\item{\texttt{Exp5} (Different model with different dataset): Similar to Exp4, but use the multilingual dataset.}
\item{\texttt{Exp6} (Different audio configuration): Use audio with 22050 Hz sampling rate, and use spectrograms with hop size 256, window size 1024.}
\end{itemize}
}

\begin{table*}[t]
\caption{Performance variation according to different parameter settings of $\varepsilon=\{\varepsilon^*{-}0.1, \varepsilon^*, \varepsilon^*{+}0.1\}$. 
The performance of unseen TTS and unseen-to-unseen VC tasks is reported in the format of WER / RECS.
VC and TTS results are compared within each.
The setting $\varepsilon{=}\varepsilon^*$ shows the best results for both VC and TTS in most of the experiments.
The best results within each experiment
are marked in bold.
}
\label{tab:epsilon_optim}

\resizebox{\linewidth}{!}{
\begin{tabular}{|c|l|c|c|c|c|c|c|c|c|}
\hline
                     \emph{(WER / RECS)} & \multicolumn{1}{c|}{}           & \begin{tabular}[c]{@{}c@{}}GT\\ (LibriTTS test)\end{tabular} & \begin{tabular}[c]{@{}c@{}}YourTTS\\ (LibriTTS test)\end{tabular} & \begin{tabular}[c]{@{}c@{}}Exp1\\ (VCTK)\end{tabular} & \begin{tabular}[c]{@{}c@{}}Exp2\\ (LibriTTS test)\end{tabular} & \begin{tabular}[c]{@{}c@{}}Exp3\\ (VCTK)\end{tabular} & \begin{tabular}[c]{@{}c@{}}Exp4\\ (VCTK)\end{tabular} & \begin{tabular}[c]{@{}c@{}}Exp5\\ (LibriTTS test)\end{tabular} & \begin{tabular}[c]{@{}c@{}}Exp6\\ (VCTK)\end{tabular} \\ \hline
$\varepsilon^*$      & \multicolumn{1}{c|}{}           & N.A                                                          & N.A                                                               & 0.25                                                  & 0.25                                                           & 0.43                                                  & 0.25                                                  & 0.25                                                           & 0.275                                                 \\ \hline
\multirow{3}{*}{Unseen-to-unseen VC}  & $\varepsilon=\varepsilon^*-0.1$ & \multirow{6}{*}{3.78 / 0.862}                                & \multirow{3}{*}{11.79 / 0.790}                                    & 21.09 / 0.720                                         & 16.99 / 0.758                                                  & \textbf{24.20 / 0.727}                                & 18.04 / 0.703                                         & 8.08 / 0.753                                                   & 13.32 / 0.719                                         \\ \cline{2-2} \cline{5-10} 
                     & $\varepsilon=\varepsilon^*$     &                                                              &                                                                   & \textbf{14.46} / 0.740                                & \textbf{5.14 / 0.811}                                          & 31.19 / \textbf{0.727}                                & \textbf{13.10 / 0.726}                                & \textbf{4.95} / 0.782                                          & \textbf{12.26} / 0.737                                \\ \cline{2-2} \cline{5-10} 
                     & $\varepsilon=\varepsilon^*+0.1$ &                                                              &                                                                   & 22.14 / \textbf{0.751}                                & 7.36 / 0.809                                                   & 54.83 / 0.700                                         & 18.81 / 0.722                                         & 7.63 / \textbf{0.788}                                          & 16.87 / \textbf{0.741}                                \\ \cline{1-2} \cline{4-10} 
\multirow{3}{*}{Unseen TTS} & $\varepsilon=\varepsilon^*-0.1$ &                                                              & \multirow{3}{*}{54.62 / 0.805}                                    & 26.47 / 0.743                                         & 58.27 / 0.775                                                  & \textbf{25.47} / 0.733                                & 54.22 / 0.715                                         & 79.71 / 0.769                                                  & 21.03 / 0.748                                         \\ \cline{2-2} \cline{5-10} 
                     & $\varepsilon=\varepsilon^*$     &                                                              &                                                                   & \textbf{19.36 / 0.762}                                & \textbf{22.15 / 0.821}                                         & 37.18 / \textbf{0.736}                                & \textbf{33.24} / 0.741                                & \textbf{48.15 / 0.810}                                         & \textbf{16.15 / 0.764}                                \\ \cline{2-2} \cline{5-10} 
                     & $\varepsilon=\varepsilon^*+0.1$ &                                                              &                                                                   & 28.30 / 0.757                                         & 24.39 / 0.815                                                  & 55.77 / 0.708                                         & 38.07 / \textbf{0.751}                                & 54.25 / 0.805                                                  & 19.53 / 0.743                                         \\ \hline
\end{tabular}
}\vspace{-3mm}
\end{table*}

\paragraph{Results}
As shown 
in \cref{tab:epsilon_optim}, $\varepsilon{=}\varepsilon^*$ shows the best trade-off 
performance compared to nearby values.
In \texttt{Exp3}, although $\varepsilon=\varepsilon^* {-} 0.1$ showed better WER / CER than $\varepsilon{=}\varepsilon^*$, the model produces noisy artifacts that clearly degrade
perceptual quality.
The samples can be found in the supplementary material.
Also,
the models trained with the multilingual dataset with $\varepsilon {=} \varepsilon^*$ (\texttt{Exp2\&5}) show 
superior results over the multilingual baseline (YourTTS in \cref{tab:epsilon_optim}).
They even show VC performance quite similar to the ground truth, in terms of WER.

\paragraph{The Agnosticity of the $\varepsilon^*$ Value}
Interestingly, despite 
different audio configurations, datasets, and models, 
similar values of $\varepsilon^*$ were effective for \texttt{Exp1,2,4$\sim$6}.
This would be a useful 
result for future work, as \emph{it even dismisses the necessity of
the preliminary task of training a separate HiFi-GAN}.

\subsection{Subjective Evaluation Results}
For the Mean Opinion Score (MOS) test, we randomly sampled 10 sentences as source speeches from the evaluation set of unseen speakers.
The target voices were also randomly sampled but selected to be different from the source.
We used 
Amazon Mechanical Turk, 
and asked 15 native English speakers to rate 
the naturalness of the synthesized speech (Nat MOS), and how similar 
the synthesized voice
and target voice 
is (Sim MOS).
For limited resources and simplicity of user study, we evaluate perceptual quality only for the VC task.
Ratings ranged
from 1 to 5 in integers, 5 being the best.

\begin{table}[t]
\caption{Mean opinion scores (MOS) for unseen-to-unseen VC.
The best results within the same training set are marked in bold.
The VITS-based models are further trained to 800k steps.}
\label{tab:overall_mos}
\resizebox{\linewidth}{!}{

\begin{tabular}{cccc}
\hline
\multirow{2}{*}{Model}  & \multirow{2}{*}{Test dataset} & \multirow{2}{*}{Framework applied} & MOS (Nat / Sim)                            \\ \cline{4-4} 
                        &                               &                                    & VC (Unseen-to-unseen)                      \\ \hline
GT                      & VCTK                          & N.A                                & 3.99 $\pm$ 0.15 / 3.67 $\pm$ 0.18          \\
GT                      & LibriTTS                      & N.A                                & 4.01 $\pm$ 0.14 / 3.63 $\pm$ 0.17          \\ \hline
C-DSVAE                 & VCTK                          & $\times$                           & 3.78 $\pm$ 0.15 / 3.37 $\pm$ 0.18 \\
YourTTS                 & VCTK                          & $\times$                           & \textbf{3.81 $\pm$ 0.16} / 3.64 $\pm$ 0.16 \\
\textit{Zero-shot VITS} & VCTK                          & $\times$                           & 3.77 $\pm$ 0.15 / \textbf{3.82 $\pm$ 0.16} \\
\textit{Zero-shot VITS} & VCTK                          & $\bigcirc$                         & 3.64 $\pm$ 0.16 / 3.71 $\pm$ 0.17 \\ \hline
YourTTS                 & LibriTTS                      & $\times$                           & 3.68 $\pm$ 0.14 / 3.19 $\pm$ 0.18 \\
YourTTS                 & LibriTTS                      & $\bigcirc$                         & 3.70 $\pm$ 0.14 / 3.26 $\pm$ 0.18 \\
\textit{Zero-shot VITS} & LibriTTS                      & $\bigcirc$                         & \textbf{3.91 $\pm$ 0.12 / 3.41 $\pm$ 0.18} \\ \hline
\end{tabular}
}\vspace{-3mm}
\end{table}

\paragraph{Results}
In \cref{tab:overall_mos},
the methods using our framework 
show 
better or comparable
perceptual naturalness and speaker similarity against 
other methods,
while having significantly better WER / CER according to Tables~\ref{tab:vctk_wer_recs} and \ref{tab:epsilon_optim}.
When using the multilingual dataset, using \textit{Zero-shot VITS} with MDMM for unseen-to-unseen VC shows Nat / Sim MOS comparable to the ground truth audios.

\subsection{Discussion}
We explain why using the $\varepsilon^*$ value obtained from HiFi-GAN is effective.
When training HiFi-GAN, the perceptual quality improved as $L_{recon}$ got smaller.
This hints that converging a VITS-based model to a smaller $L_{recon}$ induces the decoder to a state where it produces high-quality audio, and thus provides a reason not to use $\varepsilon$ bigger than $\varepsilon^*$.
However, the audio quality improvement of HiFi-GAN was not significant after $L_{recon}$ got smaller past $\varepsilon^*$.
This means that pushing the model towards a $L_{recon}$ value smaller than this does not lead to
improvement, but rather hinders the model's ability to regularize the posterior to the text conditioned prior.
This provides a reason not to use $\varepsilon$ smaller than $\varepsilon^*$.

\section{Conclusion}
In this work, we first introduced the importance of tuning the trade-off between 
the reconstruction loss and the other losses of VITS-based models.
We hypothesized that the converged reconstruction loss value $\varepsilon^*$ obtained from the preliminary vocoder task with HiFi-GAN might guarantee a sufficient level of reconstruction quality across other tasks.
To train models with a specific target reconstruction loss $\varepsilon^*$, we used MDMM to enforce the model to be trained with the constraint.
This allows us to obtain a model superior to competitive baselines without searching the balance hyper-parameter tuning.
We showed our framework is generalized well across
various scenarios of different datasets and models.
Moreover, 
with a bigger multilingual dataset, our method 
achieves a quality 
close to ground truth in zero-shot VC.
These results hint that the performance of VITS-based models can be improved by pushing towards a certain low reconstruction loss value determined by the quality of the decoder.
While we show this with the HiFi-GAN vocoder in this work, it would be an interesting future direction to investigate whether our framework can be applied to other decoders.



\bibliographystyle{IEEEtran}
\bibliography{mybib}

\end{document}